\documentclass[10pt,conference]{IEEEtran}
\IEEEoverridecommandlockouts
\RequirePackage{fix-cm}

\usepackage[T1]{fontenc}
\usepackage{times}
\usepackage{microtype} 
\usepackage{booktabs}
\usepackage{multirow} 
\usepackage{pdflscape}
\usepackage{tikz}

\usepackage{graphicx}
\usepackage{pifont}
\usepackage[export]{adjustbox}
\usepackage{adjustbox}
\usepackage[utf8]{inputenc}
\usepackage{booktabs}
\usepackage{pgfplots}
\pgfplotsset{compat=1.10}
\usepgfplotslibrary{statistics}
\usepackage[numbers,sort&compress]{natbib}
\usepackage{subfigure}
\usepackage{soul}
\usepackage{tcolorbox}
\usepackage{amsmath}
\usepackage{amsfonts}
\usepackage{paralist}
\usepackage{multirow}
\usepackage{xspace}
\usepackage{color}
\usepackage{xcolor}
\usepackage[graphicx]{realboxes}
\usepackage{ifthen}
\usepackage{url}
\usepackage{fancybox}
\usepackage{enumitem}
\usepackage{listings}
\usepackage{balance}
\usepackage{ifthen}
\usepackage{graphicx}
\usepackage{amsmath}
\usepackage[linesnumbered,boxruled]{algorithm2e}
\usepackage{algorithm2e}
\usepackage{tablefootnote}
\usepackage{comment}
\usepackage{float}
\usepackage{algorithmic}
\usepackage{dirtytalk}
\usepackage{subfigure}
\usepackage{tikz}
\usepackage[normalem]{ulem}
\usepackage{colortbl}

\newlength\WIDTHOFBAR
\newlength\MAX  

\newcommand{\RqOne}{(RQ1) What are the characteristics of images used in Stack Overflow questions?\xspace}
\newcommand{\RqTwo}{(RQ2) How are images used in questions?\xspace} 

\newcommand{\RqThree}{(RQ3) How are questions with images answered? \xspace} 

\begin{document}

\title{Understanding the Role of Images on Stack Overflow}

\makeatletter
\newcommand{\linebreakand}{%
  \end{@IEEEauthorhalign}
  \hfill\mbox{}\par
  \mbox{}\hfill\begin{@IEEEauthorhalign}
}
\makeatother

\author{\IEEEauthorblockN{Dong Wang\IEEEauthorrefmark{2}, Tao Xiao\IEEEauthorrefmark{3}, Christoph Treude\IEEEauthorrefmark{4}, Raula Gaikovina Kula\IEEEauthorrefmark{3}, Hideaki Hata\IEEEauthorrefmark{5}, Yasutaka Kamei\IEEEauthorrefmark{2}}
\IEEEauthorblockA{Kyushu University, Japan\IEEEauthorrefmark{2}\\
Nara Institute of Science and Technology, Japan\IEEEauthorrefmark{3}\\
University of Melbourne, Australia\IEEEauthorrefmark{4}\\
Shinshu University, Japan\IEEEauthorrefmark{5}\\
Email: d.wang@ait.kyushu-u.ac.jp, tao.xiao.ts2@is.naist.jp, christoph.treude@unimelb.edu.au,\\
raula-k@is.naist.jp, hata@shinshu-u.ac.jp, kamei@ait.kyushu-u.ac.jp}
}

\maketitle
\begin{abstract}
Images are increasingly being shared by software developers in diverse channels including question-and-answer forums like Stack Overflow.
Although prior work has pointed out that these images are meaningful and provide complementary information compared to their associated text, how images are used to support questions is empirically unknown.
To address this knowledge gap, in this paper we specifically conduct an empirical study to investigate (I) the characteristics of images, (II) the extent to which images are used in different question types, and (III) the role of images on receiving answers.
Our results first show that \textit{user interface} is the most common image content and \textit{undesired output} is the most frequent purpose for sharing images.
Moreover, these images essentially facilitate the understanding of 68\% of sampled questions.
Second, we find that \textit{discrepancy} questions are more relatively frequent compared to those without images, but there are no significant differences observed in description length in all types of questions.
Third, the quantitative results statistically validate that questions with images are more likely to receive accepted answers, but do not speed up the time to receive answers.
Our work demonstrates the crucial role that images play by approaching the topic from a new angle and lays the foundation for future opportunities to use images to assist in tasks like generating questions and identifying question-relatedness.
\end{abstract}



\section{Introduction}
Software developers handle a wide variety of artifacts, ranging from source code and documentation to diagrams and user interfaces. While the former two are well-suited to be represented as text, the latter ones often require visual representation in the form of images. Unsurprisingly, the number of images used in electronic developer communication such as the question-and-answer forum Stack Overflow is constantly rising. 
Stack Overflow allows users
to attach additional files along with their posted content.\footnote{\url{https://stackoverflow.blog/2010/08/18/new-image-upload-support/}}
Nayebi~\cite{2020_eye} reported a linearly increasing trend in the number of posts including an image on Stack Overflow in the past five years.

Developers often rely on Stack Overflow to search for various information on the Web.
Stack Overflow as a large knowledge base has accumulated millions of programming-related questions and answers. 
Effectively obtaining answers is essential for developers, which could further save their time and effort on software development. 
However, several studies pointed out that it is not trivial for developers to formulate good quality questions~\cite{ponzanelli2014improving, rahman2018evaluating, liu2022formulate}. 
The literature shows that a significant percentage of the questions on Stack Overflow are extremely poor in nature~\cite{correa2014chaff}.  
Consequently, these poor-quality questions are less likely to receive answers and get solved, affecting users' experience.  
For example, recent work by Yazdaninia et al.~\cite{9462981} found that 47\% of the Stack Overflow questions had not received accepted answers.

Prior work has reported that providing
information in a way that promotes readability and comprehension is regarded as the strongest indicator of the quality of the question~\cite{duijn2015quality}.
Apart from the regular text-based elements (question tags, titles, and descriptions), many studies have been carried out to understand code snippets and investigate the role of the presence of code snippets in question quality~\cite{calefato2015mining, gao2020generating}.
For example, Asaduzzaman et al.~\cite{asaduzzaman2013answering} observed that question categories that contain a code snippet have a high answer ratio and may have more than one possible good answer.
Duijn et al.~\cite{duijn2015quality} demonstrated that quality questions need quality code snippets.
However, image, another visual communication method to convey information, is not widely studied in the Stack Overflow domain.

Previous research on images on Stack Overflow~\cite{2020_eye} has been conducted, but it does not consider the role of images in asking a question and how it is answered.
To address this, we conducted an empirical study based on 1,426,980 questions that contain 2,026,090 images.
More specifically, we set out to characterise the images, investigate how these images are used in the questions, and quantitatively analyze their role in receiving answers. 
We formulate the following three research questions to guide our study:
\begin{itemize}
    \item \textbf{\RqOne}\\
    \ul{\emph{Motivation:}} The survey by Nayebi~\cite{2020_eye} stated that developers are increasingly sharing images in social coding environments (e.g., Stack Overflow). 
    However, it remains empirically unclear what characteristics these images have (e.g., source, content, and purpose).
    Analyzing their characteristics will lay the foundation for understanding the role of images as complementary pieces of information when asking questions.
    \\
    \ul{\emph{Results:}} In terms of the image source, we find that the desktop application is the most common image source (47\%), followed by the web browser (26\%). 
    In terms of image content, we observe that the images frequently display user interfaces (58\%), while they are rarely used to depict source code and configurations (5\% and 2\%, respectively).
    Three purposes for attaching images are identified, with the undesired output being the most commonly used (45\%).
    From our manual analysis, we observe that 68\% of images are essential to their surrounding Stack Overflow questions, i.e., these questions cannot be understood without the images.
    \item \textbf{\RqTwo}\\
    \ul{\emph{Motivation:}} 
    Existing work has studied the distribution of types of general Stack Overflow questions, with nine types being classified (e.g., \textit{how-to}, \textit{discrepancy}, and \textit{error})~\cite{chris_2011}.
    Considering the nature of visuals, we assume that images could be utilized in the specific question types by users.
    Answering this research question would help users gain a better understanding of the extent to which images are used in different question types.
    \\
    \ul{\emph{Results:}} The results demonstrate that \textit{how-to}, \textit{discrepancy}, and \textit{error} are the three most frequent types of questions that contain images. 
    Moreover, \textit{discrepancy} questions are relatively more frequent compared to the questions without images, with an increase of 6\%. 
    In addition, our statistical test shows that there is no significant difference in description length between questions with images and those without images, but images tend to be used by experienced users.
    \item \textbf{\RqThree}\\ 
    \ul{\emph{Motivation:}} To efficiently obtain accepted answers is beneficial for users searching for crowdsourcing knowledge on Stack Overflow~\cite{calefato2015mining, 9462981}. 
    Several works have analyzed the factor of various non-textual elements contained in questions (e.g., URLs and code snippets). 
    For example, Calefato et al.~\cite{calefato2018ask} found that the presence of code snippets is positively associated with successful questions.
    However, the effect of the image, a popularly used non-textual element in modern society, remains largely unknown. 
    Thus, we would like to explore whether or not the presence of images also plays a role in the answer outcome.
    \\
    \ul{\emph{Results:}} Our statistical results show that questions with images are significantly more likely to receive accepted answers significantly, including the common question types when compared to the questions without images. Images do not speed up the time to receive answers.
\end{itemize}

In summary, our contributions are three-fold: \textit{first}, we construct a taxonomy of image sources, contents, and purposes in the Stack Overflow domain using manual classification, which could be applied to future large-scale studies employing automatic classifiers; \textit{second}, our study sheds light on the usage of images in different question types by mapping image characteristics and question characteristics;
\textit{third}, the empirical results highlight the role of images in understanding the questions and receiving answers, complementing knowledge regarding the question quality.

The remainder of this paper is organized as follows. Section 2 describes the data preparation. 
Sections 3--5 present the experiments that we conducted
to address RQ1--RQ3 with their results, respectively.
Section 6 discusses the implications and challenges of our study.
Section 7 situates this paper with respect to related work on Stack Overflow and non-textual information sharing in software development.
Section 8 discloses threats to the validity of our study.
Finally, we conclude the paper in Section 9.
To facilitate replication and future work in the area, we have prepared a replication package online, which includes all manually labeled datasets~\cite{replication}.

\begin{figure}[]
    \centering
    \subfigure[The proportion of SO questions that have images in an interval of a year.]{\includegraphics[width=0.45\textwidth]{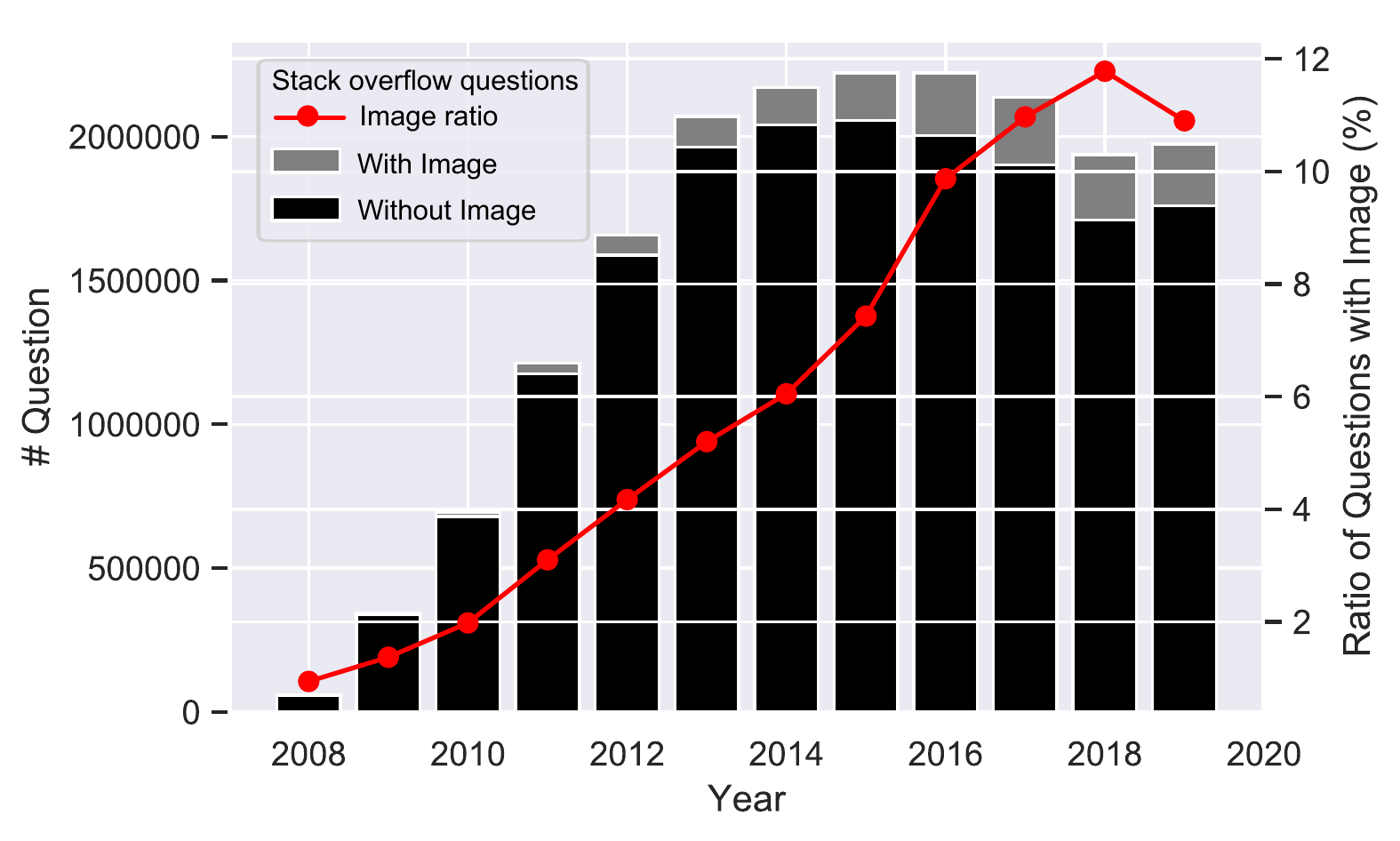}}
    \subfigure[The popular tags of SO questions that have images.]{\includegraphics[width=0.45\textwidth]{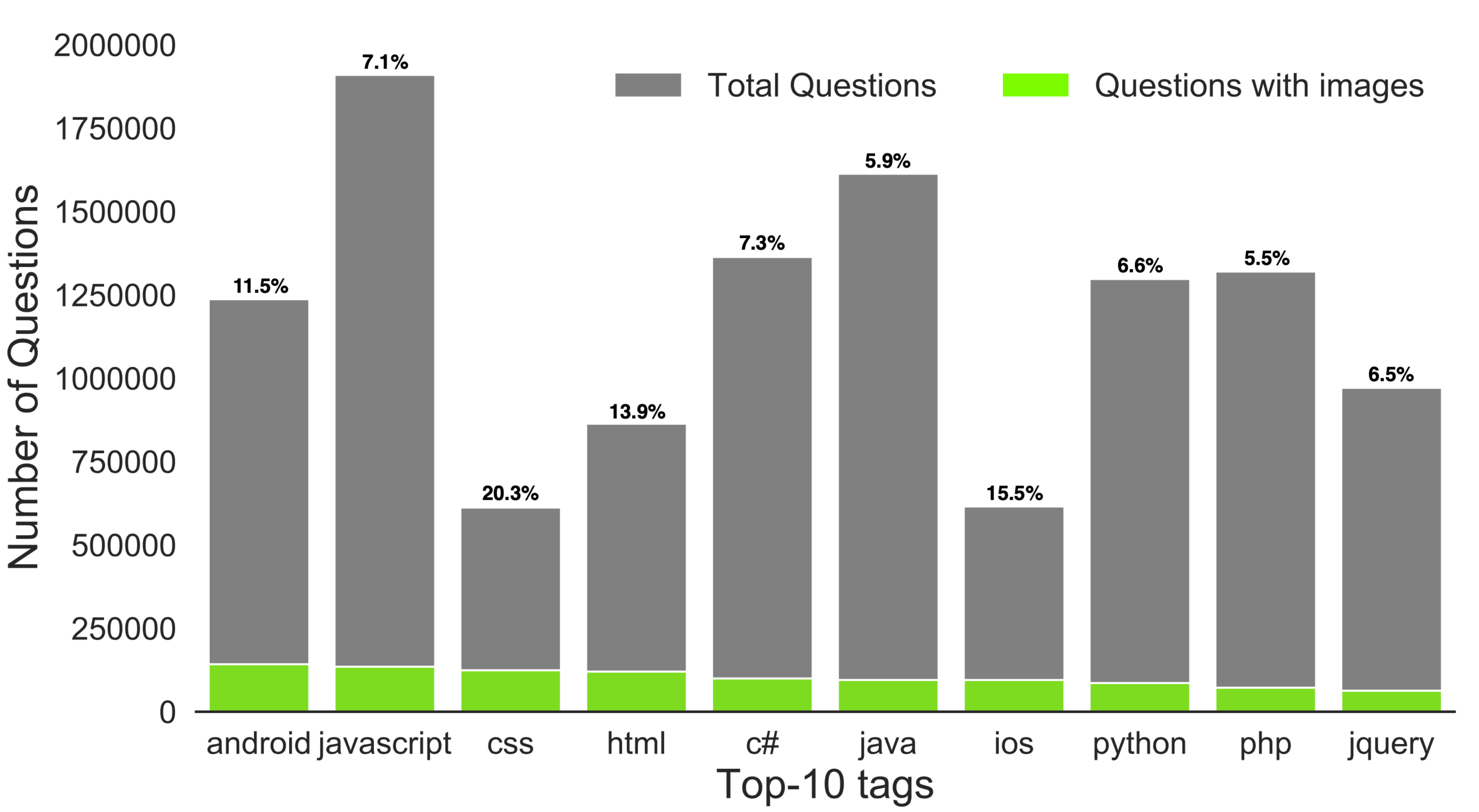}}
    \caption{Descriptive analysis of the popularity of SO questions that have images.}
    \label{fig:example}
\end{figure}

\section{Data Preparation}
In this section, we present the preparation of the data consisting of the collection of image data and the description of the dataset.

\textbf{Data collection.} To understand the role of images in Stack Overflow questions, we mined Stack Overflow using the SOTorrent dataset as of December 2019~\cite{sotorrent_2019}.
Table~\ref{tab:summary} shows a summary of our studied dataset.
The dataset consists of 18,699,426 questions. To extract questions that contain images, we use regular expressions to match hyperlinks in question bodies.
Specifically, we use a list of image file extension types (i.e., JPG, PNG, GIF, WEBP, TIFF, PSD, RAW, BMP, HEIE, INDD, JPEG, SVG, AI, EPS, and PDF) to retrieve image-related links. Using this method, we were able to obtain 1,426,890 questions that contain 2,026,090 images, which represents approximately 7\% of all questions.

\textbf{Dataset Description.} Before we address each research question, we briefly characterise the popularity of questions that contain images.
To do so, we conduct two analyses to visualize (i) the trend of questions that contain images and (ii) popular tags of the questions that contain images.
For trend analysis, we measure the proportion of questions that contain at least one image in the question body in an interval of one year.
For the tag analysis, we first extract the top 10 tags sorted by the total number of questions that contain images.
Then we measure the proportion of questions that contain at least one image for each of these top-10 tags (i.e., $\frac{Questions\_with\_image\_tag}{Total\_Question\_tag}$).

Figure \ref{fig:example} (a) shows the trend of Stack Overflow questions that contain images during our studied time span. 
The figure shows that the proportion of questions that have images out of all questions generally follows an upward trend.
Especially, from 2016, the proportion of image-based questions is steadily greater than 10\%.
Figure \ref{fig:example} (b) shows the top 10 tags that are labeled with questions with images on Stack Overflow. As shown in the figure, \textit{android} accounts for the largest number of questions that have images. 
However, in terms of proportion, the questions labeled with the \textit{css} tag most frequently contain images in the top-10, accounting for 20.3\%.

\begin{table}[]
\centering
\caption{Summary of the studied dataset}
\label{tab:summary}
\begin{tabular}{lr}
\toprule
Studied Period                                & 2008.06 -- 2019.12       \\
\# Questions                             & 18,699,426      \\
\# Questions Contain Images              & 1,426,890 (7\%) \\
\# Total Images                               & 2,026,090       \\
\# Images per Posts  (1st Qu./Median/3rd Qu.) & 1/1/2           \\ \bottomrule
\end{tabular}
\end{table}

\section{\RqOne}

\subsection{Approach} To understand the characteristics of images on Stack Overflow, we conducted a qualitative study of a statistically representative sample of all developer questions that contain at least one image in our dataset.
Since images may come from different sources and contain different types of content, we adopt three dimensions of 
(a) the image source, (b) the image content, and (c) the purpose served by the image, which is similar to prior work \cite{hata20199,tao}. 
Furthermore,  to inform the tool
design on whether support for images is crucial, we analyze the relationship between the image and the comprehension of the question understanding. 
Hence, we in addition manually classify the essentialness of the images.
We now describe the construction of the representative sample and manual coding.

\textbf{Representative sample construction.} As the entire dataset is too large to manually examine image characteristics, we draw a statistically representative sample.
The required sample size was calculated so that our conclusions
about the ratio of images with a specific characteristic would
generalize to all developer questions containing images with a confidence
level of 95\% and a confidence interval of 5. The calculation of
statistically significant sample sizes based on population size,
confidence interval, and confidence level is well established~\cite{sample_size}. 
Based on the calculation\footnote{https://www.surveysystem.com/sscalc.htm}, we randomly sample 384 developer questions that contain at least one image.

\textbf{Manual coding.} The qualitative analysis to understand image characteristics was conducted in multiple rounds following a systematic method similar to the work of Hata et al.~\cite{hata20199}.
In the first round, the first three authors independently
coded 30 images from the representative sample, discussed a common coding guide, and tested this coding guide on another 10 images, refining the guide, merging codes, and adding
codes that had been missed.
In the second round, to evaluate the level of agreement of our constructed codes, we calculated the Kappa agreement of our codes among the first three authors, who independently coded (a) the source, (b) the content, (c) the purpose, and (d) the essentialness of the images of another 30 randomly selected samples.
The free-marginal Kappa scores for the content are 0.88 or `Almost perfect' agreement, while the Kappa scores for the image source and image purpose are 0.75 and 0.71, respectively, which indicates `Substantial' agreement~\cite{viera2005understanding}.
However, the Kappa score for the essentialness of the images was not yet sufficient.
Thus, in the third round, we refined our definition through discussion and randomly selected another 20 samples to test the refined coding criteria for the essentialness of the images among the first three authors.
As a result, the Kappa score for the essentialness of the image reached 0.9 or `Almost perfect' agreement.
Based on this encouraging result, the
remaining data was divided into the three sets equally and then coded by the first three authors independently.
Note that one question may contain more than one image; however, in our study, we only manually classify the first image that appears in the post as we assume that images used in a question are likely to present one specific topic.
In other words, a total of 384 images were classified during our qualitative analysis.

\subsection{Results}

\begin{table*}[t]
\centering
\caption{The frequency of image sources, contents, and purposes (RQ1).}
\label{tab:rq1_characteristic}
\resizebox{.82\textwidth}{!}{
\begin{tabular}{llp{7cm}r@{}r}
\toprule
                         & \textbf{Category}   &  \textbf{Description/Example} & \multicolumn{2}{c}{\textbf{Frequency}}  \\ \midrule
\multirow{4}{*}{\textit{Source}}  & Desktop application   &      \url{shorturl.at/jnoHS}   &     180 & (47\%)       \\
                         & Web browser &  \url{shorturl.at/iqrGM}               &   99 & (26\%)        \\
                         & User-created   &   \url{shorturl.at/doBGR}           &   42 & (11\%)         \\
                         & Mobile application  &  \url{shorturl.at/iqrGM}               &  40 & (10\%)  \\
                         
                        \midrule
\multirow{6}{*}{\textit{Content}} & User interface  &   such as buttons or an input
screen, e.g., \url{shorturl.at/rsDPS}          &     224 & (58\%)       \\
& Diagram  & such as designs or charts, e.g., \url{shorturl.at/ksvIQ}                   &   43 & (11\%)          \\
                         & Result  &  such as output from a program, e.g., \url{shorturl.at/fiwMN}                  & 38 & (10\%)            \\
                         & Error code  &  such as a warning or error log, e.g., \url{shorturl.at/auzCU}               &   33 & (9\%)          \\
                         & Source code &  such as code snippets, e.g., \url{shorturl.at/gjzJT}               & 19 & (5\%)            \\

                         & Configuration &  such
as build or dependency files, e.g., \url{shorturl.at/kFGSU}             &  6 &(2\%)           \\ \midrule
\multirow{3}{*}{\textit{Purpose}} 
                         & Undesired output &  an image presents an output that a developer would
like to avoid, e.g., \url{shorturl.at/arFX1}      &    174 & (45\%)        \\
                         & Additional context  &  an image is used to further support understanding of the
developer question, e.g., \url{shorturl.at/akyU}              &  141 & (37\%)         \\
& Desired output  &   an image show an output that a developer would like to
achieve, e.g., \url{shorturl.at/cgmsF}         &    48 & (13\%)        \\
\midrule
\multirow{2}{*}{\textit{Essentialness}} & Essential  &   the question is not understandable without images.        &    263 & (68\%)       \\
                         & Not essential &  the question is understandable without images &    101 & (26\%)       \\
\bottomrule
\end{tabular}
}
\end{table*}

\textbf{Source, Content, and Purpose}. Four categories, six categories, and three categories are classified for the image source, content, and purpose through qualitative analysis, respectively.
For those instances that cannot fit the constructed categories, we regard them as others including the images that are not able to be accessible with 404. 

Table \ref{tab:rq1_characteristic} shows the descriptions along with real examples in the form of short URLs and the frequency results of the image characteristics studied. 
\ul{\textit{For the image source}}, we find that desktop application is the most common source, accounting for 47\%.
The next frequent source is web browser, with 26\% of the images being classified.
User-created and mobile applications are less common (11\% and 10\%, separately).
\ul{\textit{For the image content}}, our analysis shows that images most frequently present a user interface, accounting for 58\% of the classified images, followed by \textit{Diagram} (11\%).
The contents of source code and configuration are relatively less common, accounting for 5\% and 2\%, respectively.
\ul{\textit{For the image purpose}}, we find that the majority of images (45\%) are used to provide undesired output (showing an output that the developer would like to avoid), while 13\% of images are used to provide desired output (showing an output that a developer would like to achieve).
The rest of the images are served to enable additional context (image is used for further understanding of the developer post), i.e., 37\%.
In addition, with respect to the essentialness, we observe that 68\% of images are essential to the question context, which means that these questions cannot be understood without the images.
Such results indicate that the information conveyed from the images may fulfill the comprehension needs of the questions.

\textbf{Relationship between image characteristics}. Furthermore, we investigate the relationship between image content and purposes by drawing a parallel category diagram.
Parallel sets are variants of
parallel coordinates, in which the width of lines that connect sets corresponds to the frequency of their co-occurrence.
Figure~\ref{fig:RQ1-Relation} depicts the relationship between image contents and purposes.  
We observe that different image contents are likely to show the primary purposes.
For example, in terms of image contents \textit{User interface}, \textit{Result}, and \textit{Error code}, images tend to present undesired outputs (52\%, 53\%, and 97\%, respectively). 
Although in terms of image contents \textit{Source code}, \textit{Diagram}, and \textit{Configuration}, images are more likely to support context understanding (89\%, 58\%, and 100\%, respectively).

\begin{figure}[]
\includegraphics[width=.5\textwidth]{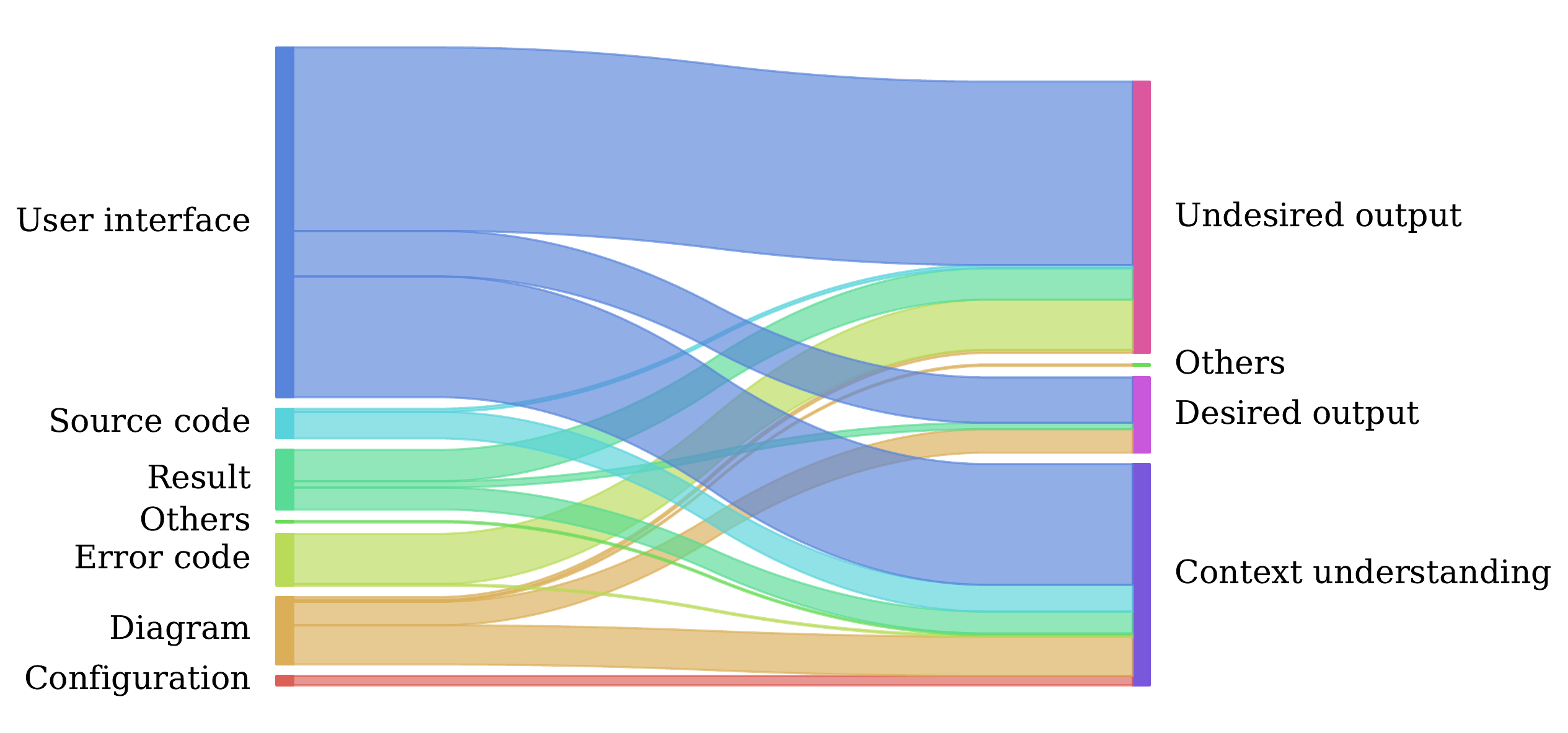}
\caption{Relationship between Image Contents and Purposes (RQ1).}
\label{fig:RQ1-Relation}
\end{figure}

\textbf{Representative Examples}.
Now we show two representative examples to illustrate the categories of the source, content, and purpose of the images.
Figure \ref{fig:representative1} (a) shows an example in which the image is captured from a desktop application called \texttt{TYPO3}. \texttt{TYPO3} is a Web content management system.
We classify the image content into user interface since the menu list and function icons of \texttt{TYPO3} are clearly displayed. 
As the text written in the question body is \textit{``But the backend content view looks ugly, see attachment image"}, the image is obviously attached to present undesired output.
The author used the word `ugly' to depict the output.
If no image is provided, the developer who attempts to answer this question would have no idea how ugly it is. 
Therefore, it is essential for the user to attach images in this case.
Figure \ref{fig:representative1} (b) shows an example of a question in which the image is taken from a mobile virtual machine. 
In this case, we classify its source as mobile application.
The mobile screen displays an interface that includes an input box; therefore, it is labeled as the user interface. 
As the question describes, the developer shares this image trying to complement the situation where the image shown can be properly placed in \texttt{android studio} while it cannot work in the specific mobile model (i.e., \texttt{Samsung Galaxy s3}).
Based on this, we classify the purpose of the image as additional context.
In this instance, the image is considered essential since it is tricky for a developer to imagine the vision of the transparent buttons and the proper layout.  


\begin{figure}[]
    \centering
    \subfigure[Visual issue report that provides an undesired output.\protect\footnotemark]{\includegraphics[width=.9\linewidth]{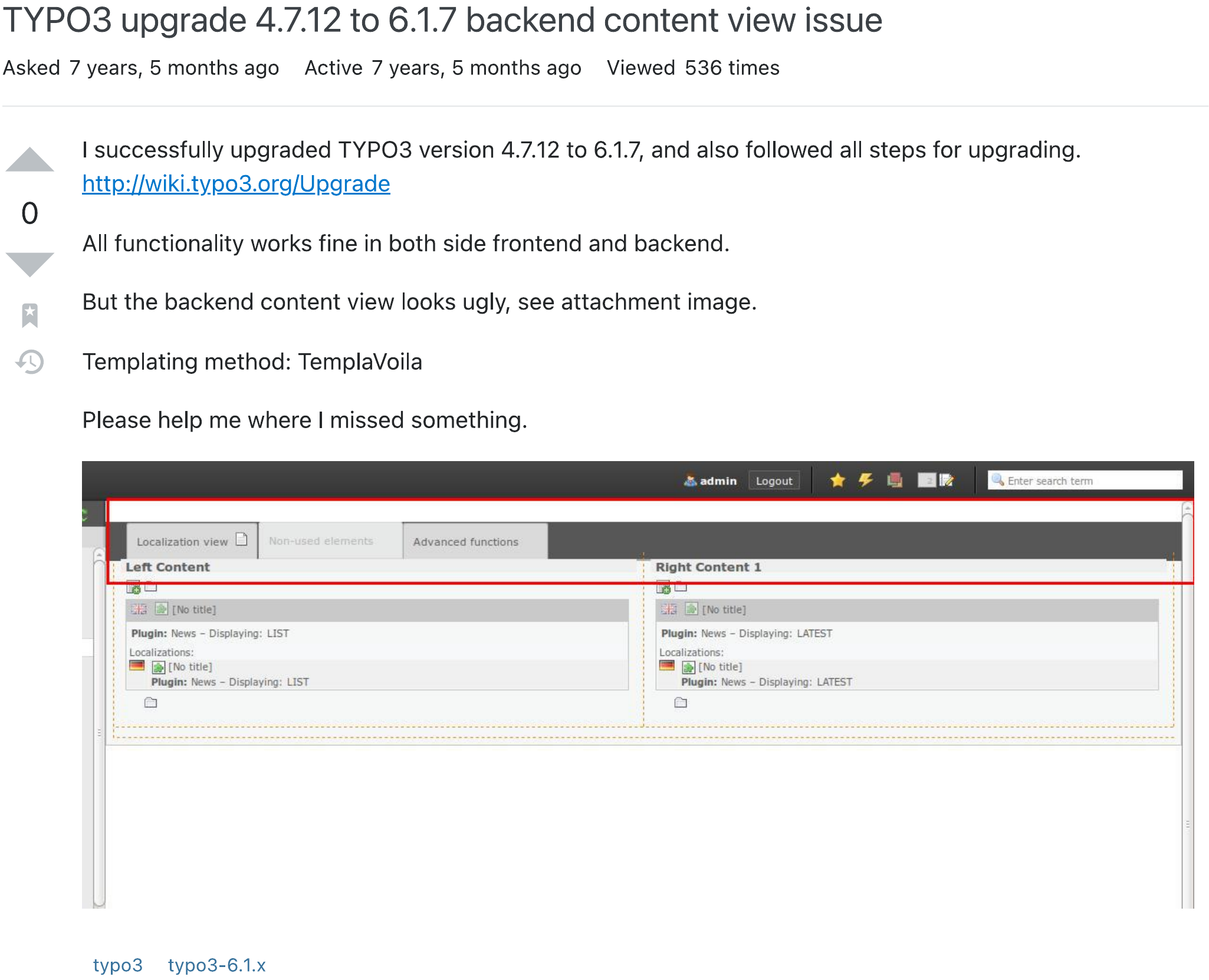}}
    \subfigure[Visual issue report that provides additional information.\protect\footnotemark]{\includegraphics[width=.9\linewidth]{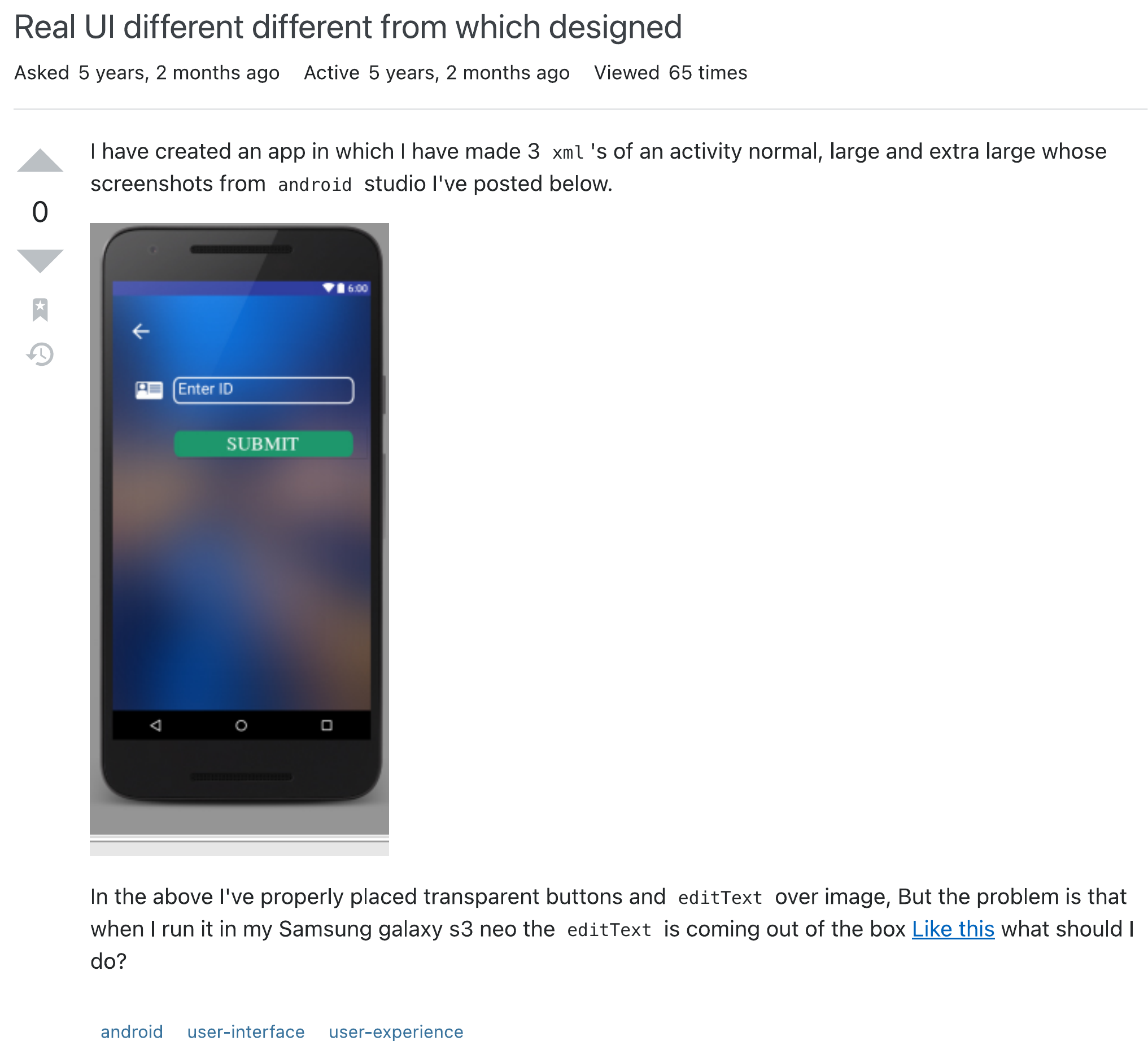}}
    \caption{Two purpose kinds of attaching visual contents.}
    \label{fig:representative1}
\end{figure}

\begin{tcolorbox}
\textbf{RQ1 Summary:} Our results show that desktop application is the most common image source (47\%), followed by the web browser (26\%). The images are more likely to contain the content of the user interface (58\%), while the contents of source code and configuration are the least common (5\% and 2\%, respectively).
We identify three types of purposes for sharing images in questions, with the undesired output being the most common (45\%).
Furthermore, we observe that 68\% of images are essential to the question context, and the questions cannot be understood without these images.
\end{tcolorbox}

\begin{figure*}[]
    \centering
    \subfigure[Question Types and Image Contents]{\includegraphics[width=0.45\textwidth]{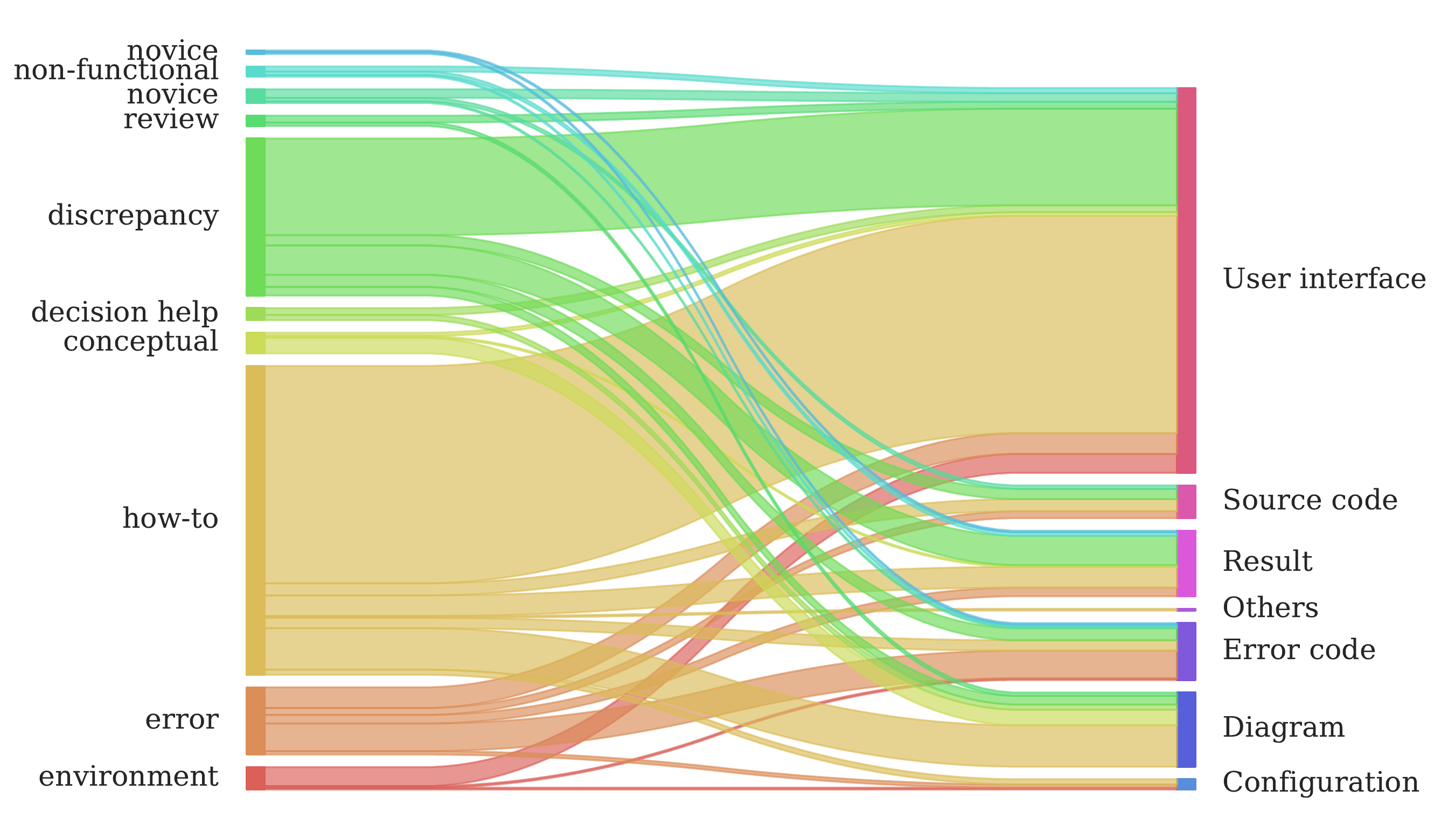}}
    \subfigure[Question Types and Image Purposes.]{\includegraphics[width=0.5\textwidth]{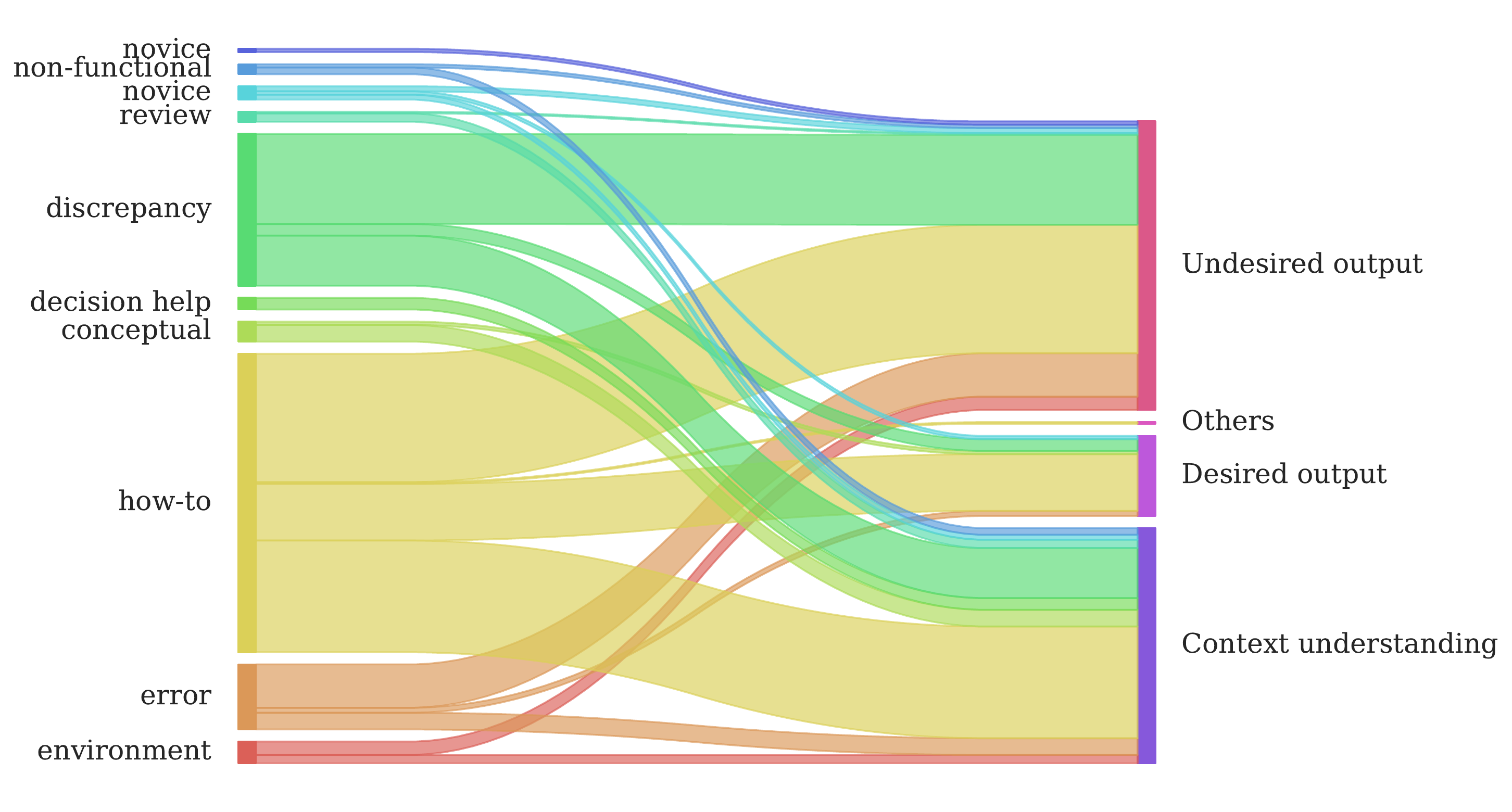}}
    \caption{Relations between Question Types, Image Contents, and Image Purposes (RQ2)}
    \label{fig:RQ2}
\end{figure*}

\footnotetext[2]{\url{https://stackoverflow.com/questions/21459190/}}
\footnotetext[3]{\url{https://stackoverflow.com/questions/36866090/}}

\section{\RqTwo}

\subsection{Approach}
To understand what types of questions are likely to include images, we perform a qualitative analysis to classify 384 representative samples that are used in RQ1.
For the types of questions, we refer to the coding schema provided by a pioneering work by Treude et al.~\cite{chris_2011}. 
The coding schema consists of the following nine categories: 
\begin{itemize}
    \item \textit{how-to}: Questions that ask for instructions.
    \item \textit{discrepancy}: Some unexpected behavior that the person asking the question wants explained.
    \item \textit{environment}: Questions that ask the environment either during development or after deployment.
    \item \textit{error}: Questions that include a specific error message.
    \item \textit{decision help}: Questions that ask for an opinion.
    \item \textit{conceptual}: Questions that are abstract and do not have a concrete use case.
    \item \textit{review}: Questions that implicitly or explicitly ask for a code review.
    \item \textit{non-functional}:  Questions about non-functional requirements such as performance or memory usage.
    \item \textit{novice}: Often explicitly states that the person asking the question is a novice.
    \item \textit{noise}: Questions not related to programming.
\end{itemize}
Similarly to RQ1, we then conducted our manual coding in several rounds following a systematic method.
Since the coding book already existed, two authors participated in this manual analysis.
We first selected 30 questions from our representative samples and the first two authors independently classified them into nine types of questions, in order to validate the comprehensive understanding of the coding schema.
However, the measured fixed-marginal Kappa score was not sufficient in the first round.
Then they opened discussions and tried to reach a consensus on disagreements between them.
During these discussions, the third author, who has more than ten years of Stack Overflow research experience, joined as a referee to resolve each disagreement. 
In the second round, the first two authors coded another 30 samples and the fixed-marginal Kappa score reached 0.90, indicating almost perfect.
Inspired by this result, the remaining representative samples (i.e., 324 questions) were equally divided into two sets, and the first two authors independently coded them, respectively.

\textbf{Control Group.} In addition, we construct a control group to fairly compare the difference in terms of question types between questions with images and those without images. 
To do so, we randomly selected an equal number of 384 questions without images.
Since the measured agreement between the first two authors was already almost perfect, similarly, we divided 384 questions without images into two sets (i.e., 192 for each set).
The first author independently coded the first set and the second author independently coded the second set.

After we classified the types of sampled questions with images and the control group, we further analyzed the question characteristics, especially the difference in \textit{(I) the description length} and \textit{(II) the experience of the questioner} between them. 
Kuramoto et al.~\cite{kuramoto2022visual} observed that issue reports with images are described in fewer words
than non-visual issue reports.
Inspired by their work, we assume that images could also reduce the number of words to generate questions.
Existing work pointed out that experienced users tend to generate better questions to maintain their reputation~\cite{ponzanelli2014improving}.
We conjecture that image as a strategy would be more often used by experienced questioners, as they may better summarize the key information from the image contents.
Therefore, we introduce the following two related hypotheses to test:

\noindent
\textit{\ding{111}~(H1) ``A question with images includes fewer words.''}\\
\textit{\ding{111}~(H2) ``A question with images is more likely to be submitted by an experienced user.''}

\noindent
Note that to calculate the description length of a question, we specifically count the number of words in the body description, excluding its title. Meanwhile, URLs to attach images, code blocks, and tables are not counted as words.
The experience of the questioner refers to the number of questions that were submitted before the studied question with images, similar to the prior work~\cite{experience}.
To validate the statistical significance of the two proposed hypotheses, we elect to adopt Bonferroni's multiple comparison test~\cite{simes1986improved} with $\alpha$= 0.05 considering multiple pairwise comparisons, 
It is a common method that can be used to compare different groups at the baseline~\cite{lee2018proper}.

\begin{table}[b]
\centering
\caption{The frequency of question types that contain images (RQ2).}
\label{tab:rq2_question}
\begin{tabular}{llrr}
\toprule
                         & \textbf{Category}   &  \textbf{With image} &  \textbf{Without image} \\ \midrule
\multirow{9}{*}{\textit{Question Types}}  & how-to    &     \cellcolor{blue!20}185 (46\%) &   162 (42\%)   \\
                         & discrepancy &    \cellcolor{blue!20}93 (24\%) &   69 (18\%)    \\
                         & error                 &  41 (10\%) & 55 (14\%)\\
                         & environment              &   13 (3\%) &    16 (4\%)     \\
                         & conceptual              &   13 (3\%) &   10 (3\%)      \\
                         & novice              &   13 (3\%) &    11 (3\%)     \\
                         & decision help             &   7 (2\%) &   25 (7\%)      \\
                         & non-functional              &   8 (2\%) &    10 (3\%)     \\
                         & review              &   7 (1\%) &  14 (4\%)       \\
                        \bottomrule
\end{tabular}
\end{table}

\subsection{Results}
\textbf{Question Types}. Table~\ref{tab:rq2_question} shows the frequency of types of questions that contain images.
As we can see, the most common type of question is \textit{how-to} (questions that ask for instructions), accounting for 46\%.
The following frequent type is \textit{discrepancy}, with 24\% of instances being classified.
Although these two most frequent types are consistent with questions without images, we find that the proportion of \textit{discrepancy} is relatively higher (24\% versus 18\% as shown in Table~\ref{tab:rq2_question}).
This finding suggests that users are more likely to upload images with the intention of receiving explanations for some unexpected behavior. 
On the other hand, we observe that the third common category \textit{error} is less likely to contain images (10\% versus 14\%).
One possible reason is that, compared to other visual images (such as icons in the user interface), the structure of the error message is organized and could be easier to be replaced using plain text.
In addition, excluding \textit{how-to}, \textit{discrepancy}, and \textit{error} question types, images are rarely used in the other six types of questions (\textit{environment},\textit{conceptual}, \textit{novice}, \textit{decision help}, \textit{non-functional}, and \textit{review}), with a frequency of less than 5\%.

\textbf{Relationship between image characteristics and question types}. Similar to RQ1, we explore the relationships between question types and image contents, question types, and image purposes.
Figure~\ref{fig:RQ2} presents these relationships via parallel category diagrams, specifically where (a) shows the relationship between question types and image contents, and (b) describes the relationship between question types and image purposes. 
Figure~\ref{fig:RQ2} (a) intuitively shows that most of the image contents used in \textit{how-to} and \textit{discrepancy} questions are the user interface, with 70\% and 56\%, respectively.
From Figure~\ref{fig:RQ2} (b), we find that images contained in \textit{discrepancy} and \textit{error} questions tend to display undesired output, i.e., 60\% and 67\%, respectively. 

\textbf{Question Characteristics}. Figure~\ref{fig:RQ2_Metric} presents the results of the comparison of the characteristics of the questions: specifically (a) for the description length and (b) for the experience of the questioner.
As shown in Figure~\ref{fig:RQ2_Metric} (a), no big differences (close median scores) are observed for the most frequent question types (i.e., \textit{how-to}, \textit{discrepancy}, and \textit{error}). 
It may indicate that the image does not reduce the number of words to describe the questions.
Furthermore, the statistical test scores $\alpha$ are greater than 0.05 for the comparison of all types (referred to \textit{All}) and the frequent ones, suggesting that there is no significant difference.
On the other hand, as shown in Figure~\ref{fig:RQ2_Metric} (b), we visually see that there exists a difference in terms of comparison of the experience of the questioner.
For example, within the category \textit{All}, the median of \textit{Without Image} is 4 while the median of \textit{With Image} is 11.
Similar findings can be observed for the most common type of questions (\textit{how-to}).
Furthermore, the statistical test confirms the significant difference, with $\alpha$ being smaller than 0.05.
We now summarize the validation of the two proposed hypotheses:

\noindent
\textit{\ding{111}~(H1) ``A question with images includes fewer words.''} is not supported.\\
\textit{\ding{111}~(H2) ``A question with images is more likely to be submitted by an experienced user.''} is supported.

\begin{tcolorbox}
\textbf{RQ2 Summary:} Our results show that \textit{how-to}, \textit{discrepancy}, and \textit{error} are the three most common question types to contain images. 
Among them, \textit{discrepancy} questions are relatively more frequent compared to the questions without images with an increase of 6\%. 
Furthermore, our statistical test shows that there is no significant difference in description length between questions with images and the ones without images, but images tend to be used by experienced users.
\end{tcolorbox}

\begin{figure}[]
    \centering
    \subfigure[Comparison of Description Length.]{\includegraphics[width=.5\textwidth]{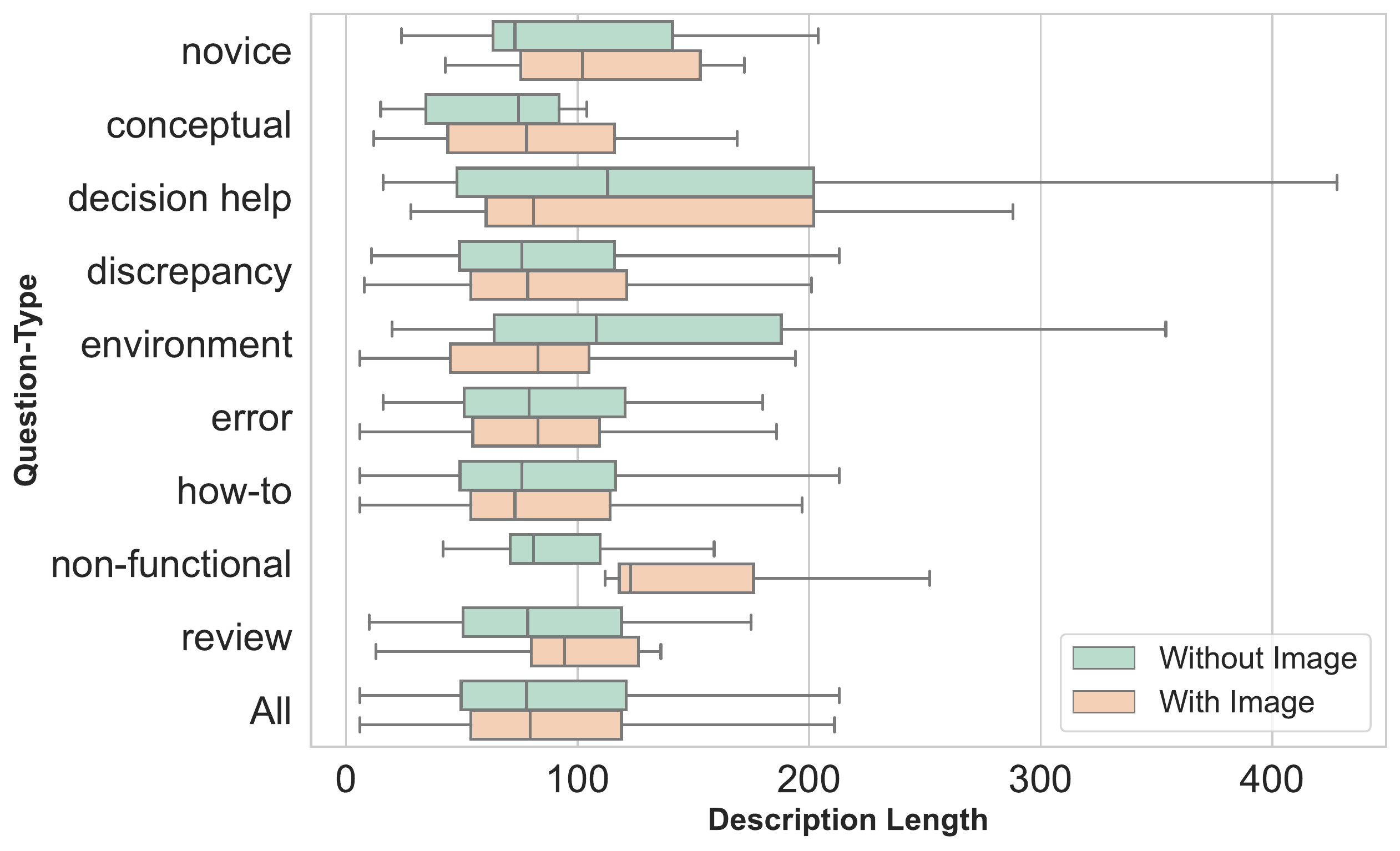}}
    \subfigure[Comparison of Questioner Experience.]{\includegraphics[width=.5\textwidth]{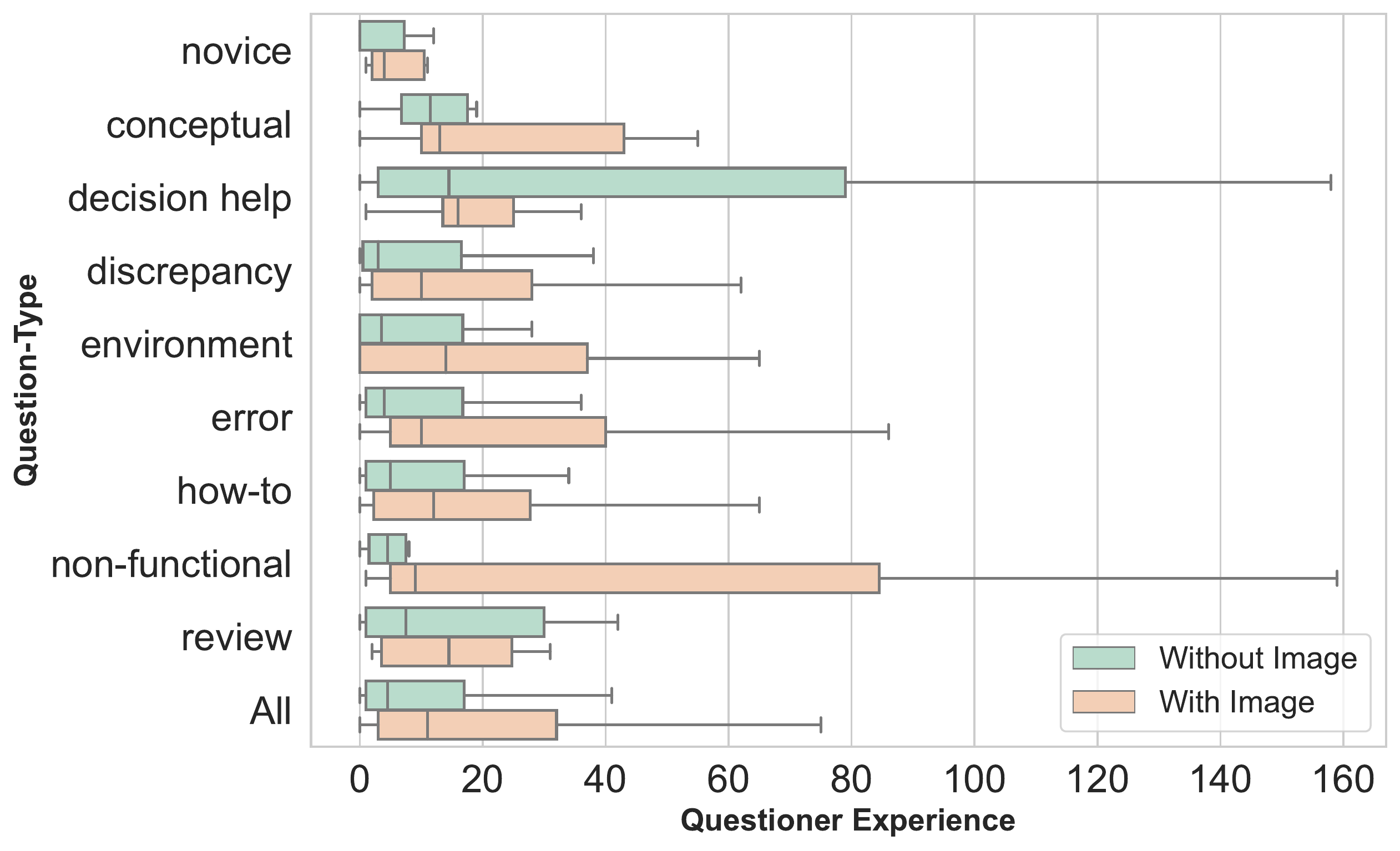}}
    \caption{Comparison of Question Characteristics (RQ2)}
    \label{fig:RQ2_Metric}
\end{figure}

\begin{table*}[]
\centering
\caption{Comparison of Answered Questions (RQ3)}
\label{fig:RQ3_answer}
\begin{tabular}{lrrrr}
\toprule
               & \multicolumn{2}{c}{Accepted Answer}              & \multicolumn{2}{c}{Has Answers}                  \\ \cmidrule{2-5} 
               & With image & Without image &  With images & Without image \\ \midrule
how-to         & \cellcolor{blue!20}110 (60\%)                 & 78 (48\%)                       & 158 (88\%)                  & 147 (91\%)                      \\
discrepancy    & 45 (48\%)                    & 32 (46\%)                      & 75 (82\%)                    & 57 (82\%)                       \\
error          & \cellcolor{blue!20}21 (51\%)                   & 21 (38\%)                      & 30 (77\%)                   & 45 (82\%)                      \\
decision help  & \cellcolor{blue!20}6 (86\%)                   & 14  (56\%)                     & 6 (86\%)                     & 23 (92\%)                      \\
novice         & 5 (45\%)                    & 5 (45\%)                       & 10 (91\%)                   & 10 (91\%)                      \\
conceptual     & \cellcolor{blue!20}9 (69\%)                    & 3 (30\%)                       & 10 (83\%)                   & 8 (80\%)                        \\
environment    & 6 (46\%)                    & 10 (63\%)                       & 11 (85\%)                   & 16 (100\%)                      \\
non-functional & 3 (50\%)                    & 4 (40\%)                       & 5 (83\%)                   & 9 (90\%)                      \\
review         & 4 (67\%)                    & 8 (57\%)                       & 6 (100\%)                    & 13 (93\%)                      \\
Total          & \cellcolor{blue!20}205 (56\%)                  & 175 (47\%)                     & 311 (85\%)                  & 328 (88\%)                      \\ \bottomrule
\end{tabular}
\end{table*}

\section{\RqThree}
\subsection{Approach}
In this research question, we aim to investigate the role of images in receiving answers.  
Specifically, we perform a quantitative method to analyze the difference between the questions with images and the ones without images in terms of the following three metrics: 
\begin{itemize}
    \item \textit{(I) Has Answers:} denotes if not a question receives an answer.
    Asaduzzaman et al.~\cite{asaduzzaman2013answering} reported that the number of unanswered questions has increased significantly.
    Their work also revealed that unanswered questions are usually not clear and too difficult to understand.
    In this sense, we assume that images could help to improve the answer rate as an image is worth a thousand words.
    \item \textit{(II) Accepted Answers:} represents whether or not a question receives an accepted answer. 
    Recent work stated that as of August 2019, 47\% of Stack Overflow questions were unresolved~\cite{9462981}.
    If developers do not get their accepted answers, the websites will not be useful. 
    Prior work~\cite{calefato2015mining} also claimed that code snippet is an important factor related to the question quality.
    Thus, similarly, we hypothesize that images may help questioners get accepted answers compared to those with only plain text.
    \item \textit{(III) Answer Time:} refers to the period between the time of submission of the question and the first answer time in this study. 
    Wang et al.~\cite{wang2018understanding} argued that the user experience when asking questions on Q\&A websites can be significantly improved by reducing the needed time to get an answer. 
    Since the image serves as complementary information, we speculate that it will help the questioners receive an answer faster.
    Note that in the computation of the time period, we exclude those questions that do not contain any answers.
\end{itemize}
Based on the three metrics and their rationales, we further propose the following three hypotheses to test:

\noindent
\textit{\ding{111}~(H3) ``A question with images receives more answers than those without images.''}\\
\textit{\ding{111}~(H4) ``A question with images receives more accepted answers than those without images.''}\\
\textit{\ding{111}~(H5) ``A question with images takes less time to receive the first answers.''}

\noindent
For the (H3) and (H4), we elect to invoke the Chi-square test, a nonparametric hypothesis test that is largely used to compare two independent proportions (i.e., between the question with image and the question without image).
For (H5), similar to RQ2, we apply Bonferroni's multiple comparison test as we have multiple categories of question types (e.g., how-to, discrepancy, etc).

\subsection{Results}
\textbf{Answer Characteristics.} 
Table~\ref{fig:RQ3_answer} presents the comparison results in terms of Has Answers and Accepted Answers and Figure~\ref{fig:RQ3_AnswerTime} shows the comparison of Answer Time.

\begin{figure}[b]
\centering
\includegraphics[width=.5\textwidth]{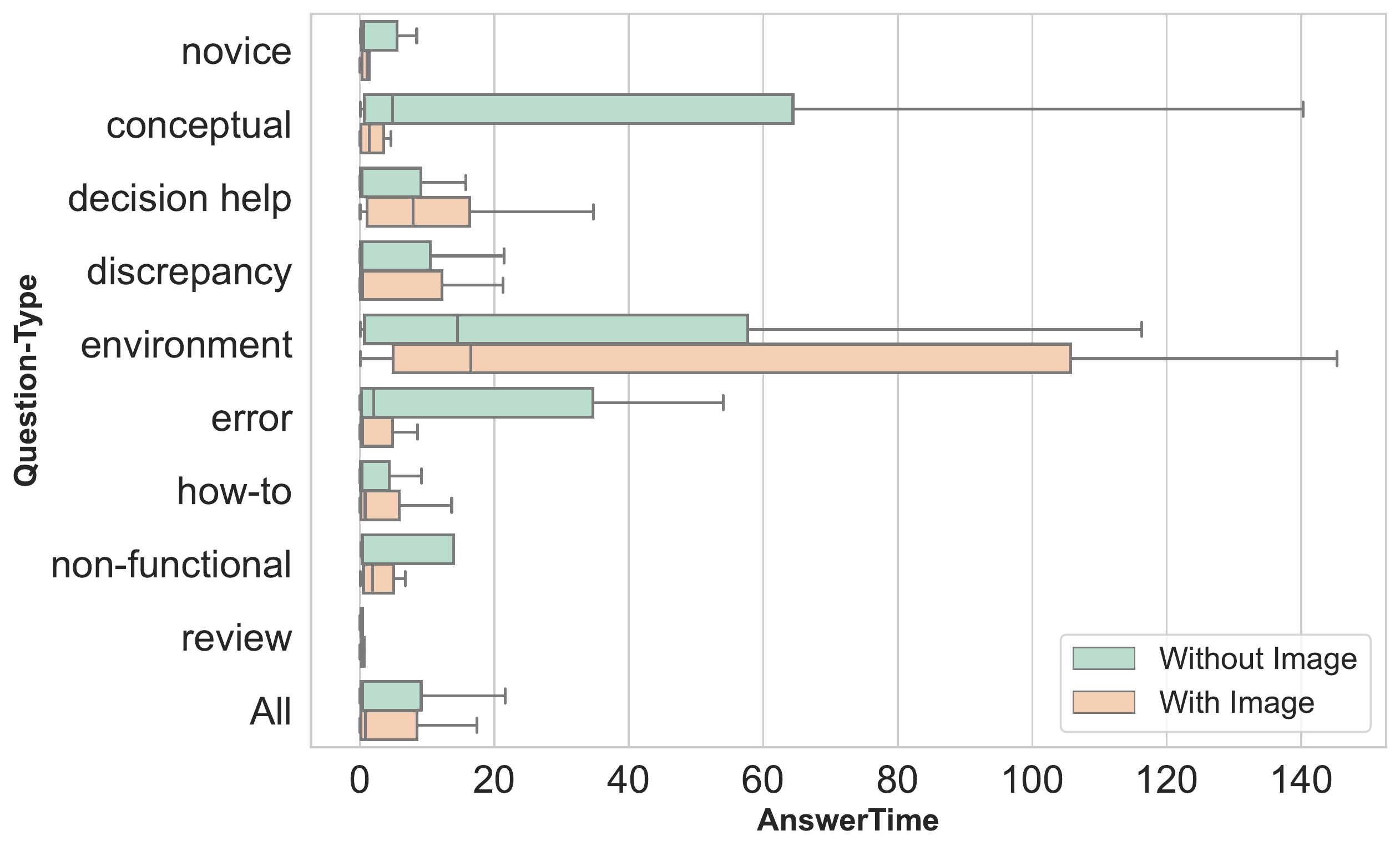}
\caption{Comparison of Answer Time (RQ3).}
\label{fig:RQ3_AnswerTime}
\end{figure}

\ul{\textit{First}}, as shown in Table~\ref{fig:RQ3_answer}, we observe that the presence of images does not greatly help to increase the probability of getting answers.
The proportion of answered questions even slightly decreases for the two common question types (i.e., \textit{how-to} and \textit{error}). 
Furthermore, the Chi-square test confirms that there is no significant difference.
\ul{\textit{Second}}, when we inspect the proportion of accepted answers, we find that, in general, questions with images are more likely to get accepted answers. 
For example, the proportion of accepted answers for questions with images is 56\%, while the proportion of accepted answers for questions without images is 47\%.
With a closer look at the types of questions, the same findings are observed within the several types including the most popular type (\textit{how-to}, 60\% against 48\%). 
The statistical tests validate that there is a significant difference in the proportion of accepted answers with $\alpha$ $<$ 0.05.
However, for the second common type (\textit{discrepancy questions}), no significant difference is found between the two groups. 
\ul{\textit{Third}}, on the contrary, no obvious difference is seen in Figure~\ref{fig:RQ3_AnswerTime} in terms of the answer time regardless of \textit{All} or the frequent question types. 
Bonferroni's test further confirms that there is no statistically significant difference.
This indicates that images may not speed up the time to get an answer.  
However, we cannot explain the causality in the current stage, as accepted answers and time could be affected by multiple confounding factors.
In the end, we summarize the validation of the three proposed hypotheses:

\noindent
\textit{\ding{111}~(H3) ``A question with images receives more answers than those without images''} is not supported.\\
\textit{\ding{111}~(H4) ``A question with images receives more accepted answers than those without images''} is supported.\\
\textit{\ding{111}~(H5) ``A question with images takes less time to receive the first answers.''} is not supported.

\begin{tcolorbox}
\textbf{RQ3 Summary:} The results show that questions with images are more likely to get accepted answers, especially the common question types (i.e., \textit{how-to} and \textit{error}).
However, the images do not significantly speed up the time to receive the first answer compared to those without images.
\end{tcolorbox}



\section{Implications and Challenges}
Based on our results for the three research questions, we now discuss the implications and challenges for both developers and researchers.

\textit{Classifying images by content, purpose, and sources provides richer information to complement a Stack Overflow question.}
Results from RQ1 reveal six types of image contents (i.e., User interface, Source code, Error code, Diagram, Results, and Configuration) and four types of image sources (i.e., Desktop Application, Web Browser, Mobile Application, and User-created).
Furthermore, we observe that images are essential and provide complementary information compared to their associated text.
As shown in Table 2, we find that 68\% of images are essential for the questions.
For example, in this question\footnote{\url{https://stackoverflow.com/questions/21631763/}}, the developer faced a problem when using Bootstrap input groups and applying the negative margin.
To seek answers, the developer posted an image showing the unexpected user interface where the bottom highlight of the input was hidden to support the problem description.
Complementing the work of Nayebi~\cite{2020_eye} who reported that 87.8\% of survey respondents agreed that images provide additional information compared to text, we empirically show how images are used and how essential they are to understand a question.
\ul{For developers, our results provide concrete recommendations on the types of images that are commonly shared in specific situations.}
Stack Overflow in general intends to be against the usage of images as raised in this post\footnote{https://meta.stackoverflow.com/q/285551}, hence our study would be used to complement the evidence-based guidelines.
For example, developers are commonly required to take a screenshot of their desktop (47\%) to present a user interface (58\%).
Furthermore, users tend to provide images when asking about undesired outputs in their user interface.
Meanwhile, in the aforementioned developer post, the developer expressed one of the major concerns of ``links to images fail''. But our classification results may also indicate that the failure cases (questions with 404 images) are relatively less frequent.
\ul{For researchers, the automatic classifiers can be proposed based on our constructed taxonomy of image characteristics for a large-scale study.} 
For example, to analyze the effect of confounding factors in future work, statistical models require quite large data, where a reliable classifier can be employed to automatically label the factors related to image features (such as contents and purposes).

\textit{
Specific questions with images would have a higher chance to get accepted answers.}
In RQ3, our results show that a significant difference in the proportion of accepted answers can be found between the questions with images and those without images. 
For example, the probabilities of \textit{how-to}, \textit{error}, and \textit{decision help} questions to receive accepted answers reach 60\%, 51\%, and 86\%, respectively, which is much higher than those with images. 
Combined with the observation in RQ2, where despite the presence of images, the description length does not decrease significantly, it may indicate that these types of questions with images would be more clear and ease comprehension. 
\ul{Therefore, we recommend that users include images and explain them properly when submitting the above specific question types.}
Prior work~\cite{ponzanelli2014improving} reported that code snippets are important for the quality of a question, but not that important for the speed of getting an answer.
Similarly, though images are an important element, in our work, there are no significant differences in terms of the answer time.
\ul{The next logical step would take into account other confounding factors and see the role of images in the statistical models.
Another possible step is to conduct a developer survey to understand how they regard questions with images. 
We hypothesize that images may potentially increase cognitive load.}

\textit{Effect of images on the Stack Overflow downstream task.} 
During the manual classification process in RQ1, we additionally annotated to what extent the images contain text. 
The results show that almost 93\% of the images studied contain text.
We conjuncture that images, which are regarded as one of the most efficient communication methods in modern society, these texts along with image features would be beneficial for the downstream work. 
Especially today, the technology of computer vision develops dramatically. 
For example, one of the state-of-the-art OCR (Optical Character Recognition) techniques named \textit{paddleocr}\footnote{https://github.com/PaddlePaddle/PaddleOCR} has played an important role in both academia and industry.
Two prospective downstream tasks are foreseen. 
For example, automatic question generation has attracted significant interest in the Stack Overflow community~\cite{duan2017question, gao2020generating}. 
The latest work specifically proposed an approach (TaskKG4Q) to generate how-to questions~\cite {liu2022formulate}. 
Another useful task is concerning question relatedness in order to relieve inefficient
and time-consuming question search~\cite{xu2018prediction, pei2021attention}.
\ul{We notice that these downstream works have not yet taken the information from the visuals (such as image contents, purposes, and text in the images). 
Thus, in future work, it is essential to evaluate whether visuals can facilitate these automatic tasks.}  

\section{Related Work}
In this section, we position our work with respect to the literature on Stack Overflow and non-textual information sharing in software development.

\subsection{Stack Overflow}
Stack Overflow is a gold mine for software engineering research and has been found to be useful for software development.
Regarding knowledge extraction from Stack Overflow, Treude et al.~\cite{chris_2011} categorized the types of questions asked and explored which questions are answered well and which remain unanswered.
Wang et al.~\cite{wang_2013} used Latent Dirichlet Allocation (LDA) topic modeling to categorise Stack Overflow questions so that they can automatically categorise new questions.
Nasehi et al.~\cite{what_make} identified characteristics of effective examples by analyzing well-received answers.
Similarly, Treude and Robillard~\cite{treude2017understanding} conducted a survey-based study to understand developers’ information needs as they relate to code fragments.
Stack Overflow is also used to understand and address challenges in new domains.
Wan et al.~\cite{wan_tse_2021} used Stack Overflow
to understand the challenges and needs among blockchain developers by applying Balanced LDA.
Abdellatif et al.~\cite{abdellatif2020challenges} examined Stack Overflow to provide insights on topics that chatbot developers are interested in and the challenges they face.
Yang et al.~\cite{yang2016security} conducted a large-scale study on Stack Overflow to
identify security-related questions asked by practitioners and investigated the popularity
and difficulty of different topics.
Apart from a large body of existing empirical work, many tools have been proposed to assist developers with the usage of Stack Overflow.
Treude and Robillard~\cite{chris_2016_icse} presented SISE, a novel machine learning-based approach, to extract insight sentences from Stack Overflow and use them to improve API documentation.
Gao et al.~\cite{Tosem_2020_xin} proposed a novel question generation task based on a sequence-to-sequence learning
approach to formulating high-quality question titles from given code snippets.
Most recently, Nayebi~\cite{2020_eye} reported a steady increase in sharing images over the past five years in Stack Overflow. 
Their developer survey showed that images are meaningful and provide complementary information compared to their associated text.

Although the textual contents and the code snippets of Stack Overflow are widely analyzed, the characteristics of images that are shared in questions remain unclear.
Thus, in this study, we conduct an empirical study to explore image characteristics and investigate their role in questions, in order to complement the literature knowledge in terms of the question quality.

\subsection{Non-Textual Information Sharing in Software Development}







Modern development tools are often integrated
with or supplemented by communication channels and
social media, to support developers’ collaboration and communication needs~\cite{chui2012social}.
Specifically, developers tend to share links, images, or videos to fulfill information needs during software development.
Several studies have shown that link sharing plays a significant role.
Hata et al.~\cite{hata20199} investigated the characteristics of links in source code comments, the purpose of these links, and how they evolve and decay, using GitHub as a data source.
Their results show that links are prevalent in source code repositories and that licenses, software homepages, and specifications are common types of link targets.
Specifically, links are rarely updated, but many link
targets evolve.
In the context of code review settings, Wang et al.~\cite{WANG_emse} observed seven intentions behind link sharing, and the results of their developer survey suggest that link sharing is useful.
Meanwhile, Hirao et al.~\cite{hirao2019fse} categorized five types of review linkage, such as patch dependency, broader context, and alternative solution, and suggested that review linkages could improve reviewer recommendation.
In addition to link sharing, developers are increasingly making use of screenshots and screen-recordings as a means of reporting problems~\cite{screencast_emse, tango_icse, WANG2019139}.
Wang et al.~\cite{WANG_emse} proposed SETU which combines information from screenshots and textual descriptions to detect duplicate crowdtesting reports.
The evaluation results showed that SETU can significantly outperform existing state-of-the-art approaches.
Cooper et al.~\cite{tango_icse} introduced TANGO to help developers determine whether video-based bug reports represent the same bug.
TANGO combines tailored computer vision techniques, optical character recognition, and text retrieval and could reduce  developer effort by more than 60\%.
The latest work~\cite{taesiri2022clip} has successfully
extracted relevant information automatically from gameplay videos to detect bugs in video games and shows promising evaluation results.

Similarly to previous studies, our work serves as the first step towards understanding the role of shared images in Stack Overflow that play in fulfilling information needs. 
Our work would further encourage future studies on the effect of images on their application and also complements other research that looks into the usage of alternative media for communications among developers (e.g., visuals, links, etc.).

\section{Threats to Validity}
We now discuss the threats to the validity of our study.

\textbf{External Validity.}
External Validity is  concerned  with our  ability  to  generalize  based  on  our  results.
We only conduct an empirical study on Stack Overflow.
As such, observations based on this study may not generalize to other Q\&A forums.
However, Stack Overflow is nowadays regarded as the most popular Q\&A forum involved with a large number of questions and developers.
On the other hand, our sample of images is diverse, including images related to varying programming languages.
Another threat may be related to the time span of the studied SOTorrent datasets (as of December 2019).
However, we believe that our proposed research questions (characteristics of images, questions, and answers) are not largely affected by the timing factor.
However, replication studies may help improve the strength of generalizations that can be drawn.

\textbf{Construct Validity.} Construct validity refers to the relation between theory and observation.
We summarize three potential threats to construct validity. 
First, in the identification of images that are shared in the questions, we rely on a list of file extension types of images.
However, cases might occur where some images are not retrieved by any type in this list.
Nevertheless, to mitigate this threat, we make sure that we cover the most common extension types, i.e., JPG, PNG, GIF, WEBP, TIFF, PSD, RAW, BMP, HEIE, INDD, JPEG, SVG, AI, EPS, and PDF.
Second, when classifying the image characteristics, we only take into account the first image appearing in the question.
However, a question may contain more than one image.
We believe that it may not have a large impact on our findings in this paper since one question tends to focus on one specific topic.
Third, in our qualitative analysis of image characteristics (i.e., source, content, purpose), and the manual annotation of question types may be miscoded due to the subjective nature of our coding approach. 
To mitigate this threat, similar to prior work~\cite{WANG_emse, wang_IST}, we took a systematic approach to test our comprehension with 30 continuous samples using Kappa agreement scores from three or two individuals (RQ1 and RQ2, separately). 
If the scores are not acceptable, we continue to have a discussion to refine and use another group of samples to evaluate until the Kappa score indicates that the agreement is substantial (0.61–0.80) or almost perfect (0.81–1.00).
Afterward, we were able to complete the coding of the rest of the sample dataset independently.

\textbf{Internal Validity.}
Internal validity is the approximate truth about inferences about cause-effect or causal relationships. 
Two related threats are summarized.
The first threat is related to the metrics that we selected and measured to understand the role of images.
For example, when we calculated the experience of users, the historical questions before the visual questions studied may also contain images, which would affect our final conclusion.
Meanwhile, apart from the presence of images, several confounding factors may have an effect on the answer time, but our goal in this stage is to shed light on the role of images.
The second threat concerns the choice of elected statistical tests.
To validate our five proposed hypotheses, we adopted different hypothesis tests (Bonferroni's multiple comparison test and chi-square test) taking into account the characteristics of the target data (whether independent or not).
The conclusion of significance may differ from the statistical tests; however, we are confident since these two tests are broadly used in empirical studies.

\section{Conclusion and Future Work}

In this paper, we perform an empirical study on Stack Overflow, to understand the role of images.
Specifically, we propose three research questions to investigate the characteristics of the image, the use of images in questions, and the role of images in receiving answers. 
Our RQ1 results primarily show that various images are shared in Stack Overflow with the undesired output being the most frequent purpose, and images are essential to understanding many questions.
RQ2 results show that \textit{how-to}, \textit{discrepancy}, and \textit{error} are the three most common types of questions that contain images, but there are no significant differences in description length.
RQ3 results show that questions with images are more likely to get accepted answers than those without images, but images do not shorten the time to receive answers.

In summary, our study highlights the role that images play in Stack Overflow questions. Specifically, the image is served as an important non-textual resource to complement the users' information needs to better understand the question context. 
The next logical step would be a deeper
study of investigating the causality between images and answers, and understanding whether images would increase the cognitive load by a user survey. 
Future research direction also includes studying the role of the images in answers, and the effect of images on downstream tasks by injecting image-related attributes.

\section{Acknowledgments}
This work has been supported by JSPS KAKENHI Grant Numbers (JP20H05706, JP21H04877, JP22K18630) and JST PRESTO Grant Number JPMJPR22P6.

\balance
\bibliographystyle{IEEEtran}
\bibliography{filteredref.bib}
\end{document}